\documentclass[twocolumn,secnumarabic,amssymb, nobibnotes, superscriptaddress, aps, prc]{revtex4-2}

\usepackage{lineno}
\usepackage[bookmarks=false]{hyperref}
\usepackage{amssymb}
\usepackage{color}
\usepackage{graphicx}% Include figure files
\usepackage{dcolumn}% Align table columns on decimal point
\usepackage{bm}% bold math
\usepackage{multirow}% bold math
\usepackage{amsmath}
\usepackage{longtable}
\usepackage{upgreek}
\usepackage{adjustbox}
\usepackage[T1]{fontenc}

\begin{document}

\title{Persistence of the $\boldsymbol{Z=28}$ shell gap in $\boldsymbol{A=75}$ isobars: Identification of a possible $\boldsymbol{(1/2^-)}$ $\boldsymbol{\upmu}$s isomer in $\boldsymbol{^{75}}$Co and $\boldsymbol{\beta}$ decay to $\boldsymbol{^{75}}$Ni}%

\author{S.~Escrig}
\email{Samuel.Escrig@csic.es}
\affiliation{Instituto de F\'isica Corpuscular, CSIC-Universitat de Val\`encia, E-46071 Val\`encia, Spain}
\affiliation{Instituto de Estructura de la Materia, CSIC, E-28006 Madrid, Spain}

\author{A. I.~Morales}
\email{Ana.Morales@ific.uv.es}
\affiliation{Instituto de F\'isica Corpuscular, CSIC-Universitat de Val\`encia, E-46071 Val\`encia, Spain}

\author{S.~Nishimura}
\affiliation{RIKEN Nishina Center, 2-1 Hirosawa, Wako, 351-0198 Saitama, Japan}

\author{M.~Niikura}
\affiliation{Department of Physics, University of Tokyo, Hongo 7-3-1, Bunkyo-ku, 113-0033 Tokyo, Japan}

\author{A.~Poves}
\affiliation{Departamento de F\'isica Te\'orica, Universidad Aut\'onoma de Madrid, E-28049 Madrid, Spain}
\affiliation{Instituto de F\'isica Te\'orica, UAM-CSIC, Universidad Aut\'onoma de Madrid, E-28049 Madrid, Spain}

\author{Z. Y.~Xu}
\affiliation{Department of Physics, University of Tokyo, Hongo 7-3-1, Bunkyo-ku, 113-0033 Tokyo, Japan}
\affiliation{RIKEN Nishina Center, 2-1 Hirosawa, Wako, 351-0198 Saitama, Japan}

\author{G.~Lorusso}
\affiliation{RIKEN Nishina Center, 2-1 Hirosawa, Wako, 351-0198 Saitama, Japan}

\author{F.~Browne}
\affiliation{School of Computing, Engineering and Mathematics, University of Brighton, Brighton BN2 4GJ, United Kingdom}
\affiliation{RIKEN Nishina Center, 2-1 Hirosawa, Wako, 351-0198 Saitama, Japan}

\author{P.~Doornenbal}
\affiliation{RIKEN Nishina Center, 2-1 Hirosawa, Wako, 351-0198 Saitama, Japan}

\author{G.~Gey}
\affiliation{LPSC, Universit\'e Grenoble-Alpes, CNRS/IN2P3, F-38026 Grenoble Cedex, France}
\affiliation{ILL, F-38042 Grenoble Cedex, France}
\affiliation{RIKEN Nishina Center, 2-1 Hirosawa, Wako, 351-0198 Saitama, Japan}

\author{H.-S.~Jung}
\affiliation{Department of Physics, University of Notre Dame, Notre Dame, Indiana 46556, USA}

\author{Z.~Li}
\affiliation{Department of Physics, Peking University, 100871 Beijing, China}

\author{P.-A.~S\"oderstr\"om}
\affiliation{RIKEN Nishina Center, 2-1 Hirosawa, Wako, 351-0198 Saitama, Japan}
\affiliation{Extreme Light Infrastructure-Nuclear Physics (ELI-NP) and Horia Hulubei National Institute for Physics and Nuclear Engineering (IFIN-HH), Reactorului 30, 077125 Bucharest-M\u{a}gurele, Romania}

\author{T.~Sumikama}
\affiliation{Department of Physics, Tohoku University, 6-3 Aramaki-Aoba, Aoba, Sendai, 980-8578 Miyagi, Japan}

\author{J.~Taprogge}
\affiliation{Departamento de F\'isica Te\'orica, Universidad Aut\'onoma de Madrid, E-28049 Madrid, Spain}
\affiliation{Instituto de Estructura de la Materia, CSIC, E-28006 Madrid, Spain}
\affiliation{RIKEN Nishina Center, 2-1 Hirosawa, Wako, 351-0198 Saitama, Japan}

\author{Zs.~Vajta}
\affiliation{Institute for Nuclear Research (Atomki), P.O. Box 51, H-4001 Debrecen, Hungary}
\affiliation{RIKEN Nishina Center, 2-1 Hirosawa, Wako, 351-0198 Saitama, Japan}

\author{H.~Watanabe}
\affiliation{IRCNPC, School of Physics and Nuclear Energy Engineering, Beihang University, 100191 Beijing, China}

\author{J.~Wu}
\affiliation{Department of Physics, Peking University, 100871 Beijing, China}
\affiliation{RIKEN Nishina Center, 2-1 Hirosawa, Wako, 351-0198 Saitama, Japan}

\author{A.~Yagi}
\affiliation{Department of Physics, Osaka University, Machikaneyama 1-1, Toyonaka, 560-0043 Osaka, Japan}

\author{K.~Yoshinaga}
\affiliation{Department of Physics, Tokyo University of Science, 2641 Yamazaki, Noda, 278-8510 Chiba, Japan}

\author{H.~Baba}
\affiliation{RIKEN Nishina Center, 2-1 Hirosawa, Wako, 351-0198 Saitama, Japan}

\author{S.~Franchoo}
\affiliation{Universit\'e Paris-Saclay, CNRS/IN2P3, IJCLab, F-91405 Orsay, France}

\author{T.~Isobe}
\affiliation{RIKEN Nishina Center, 2-1 Hirosawa, Wako, 351-0198 Saitama, Japan}

\author{P. R.~John}
\affiliation{Dipartimento di Fisica e Astronomia, Universit\'a di Padova and INFN Sezione di Padova, I-35131 Padova, Italy}
\affiliation{Institut f\"ur Kernphysik, Technische Universit\"at Darmstadt,
Schlossgartenstr. 9, D-64289 Darmstadt, Germany}

\author{I.~Kojouharov}
\affiliation{GSI Helmholtzzentrum f\"ur Schwerionenforschung GmbH, D-64291 Darmstadt, Germany}

\author{S.~Kubono}
\affiliation{RIKEN Nishina Center, 2-1 Hirosawa, Wako, 351-0198 Saitama, Japan}

\author{N.~Kurz}
\affiliation{GSI Helmholtzzentrum f\"ur Schwerionenforschung GmbH, D-64291 Darmstadt, Germany}

\author{I.~Matea}
\affiliation{Universit\'e Paris-Saclay, CNRS/IN2P3, IJCLab, F-91405 Orsay, France}

\author{K.~Matsui}
\affiliation{Department of Physics, University of Tokyo, Hongo 7-3-1, Bunkyo-ku, 113-0033 Tokyo, Japan}

\author{D.~Mengoni}
\affiliation{Dipartimento di Fisica e Astronomia, Universit\'a di Padova and INFN Sezione di Padova, I-35131 Padova, Italy}

\author{P.~Morfouace}
\affiliation{Universit\'e Paris-Saclay, CNRS/IN2P3, IJCLab, F-91405 Orsay, France}

\author{D. R.~Napoli}
\affiliation{Istituto Nazionale di Fisica Nucleare, Laboratori Nazionali di Legnaro, I-35020 Legnaro, Italy}

\author{F.~Naqvi}
\affiliation{Wright Nuclear Structure Laboratory, Yale University, New Haven, Connecticut 06511, USA}

\author{H.~Nishibata}
\affiliation{Department of Physics, Osaka University, Machikaneyama 1-1, Toyonaka, 560-0043 Osaka, Japan}

\author{A.~Odahara}
\affiliation{Department of Physics, Osaka University, Machikaneyama 1-1, Toyonaka, 560-0043 Osaka, Japan}

\author{E.~Sahin}
\affiliation{Department of Physics, University of Oslo, NO-0316 Oslo, Norway}

\author{H.~Sakurai}
\affiliation{Department of Physics, University of Tokyo, Hongo 7-3-1, Bunkyo-ku, 113-0033 Tokyo, Japan}
\affiliation{RIKEN Nishina Center, 2-1 Hirosawa, Wako, 351-0198 Saitama, Japan}

\author{H.~Schaffner}
\affiliation{GSI Helmholtzzentrum f\"ur Schwerionenforschung GmbH, D-64291 Darmstadt, Germany}

\author{I. G.~Stefan}
\affiliation{Universit\'e Paris-Saclay, CNRS/IN2P3, IJCLab, F-91405 Orsay, France}

\author{D.~Suzuki}
\affiliation{Universit\'e Paris-Saclay, CNRS/IN2P3, IJCLab, F-91405 Orsay, France}
\affiliation{RIKEN Nishina Center, 2-1 Hirosawa, Wako, 351-0198 Saitama, Japan}

\author{R.~Taniuchi}
\affiliation{Department of Physics, University of Tokyo, Hongo 7-3-1, Bunkyo-ku, 113-0033 Tokyo, Japan}

\author{V.~Werner}
\affiliation{Wright Nuclear Structure Laboratory, Yale University, New Haven, 06511 Connecticut, USA}
\affiliation{Institut f\"ur Kernphysik, Technische Universit\"at Darmstadt,
Schlossgartenstr. 9, D-64289 Darmstadt, Germany}

\author{D.~Sohler}
\affiliation{Institute for Nuclear Research (Atomki), P.O. Box 51, H-4001 Debrecen, Hungary}

\date{\today}%

\begin{abstract}

\textbf{Background:} The evolution of shell structure around doubly magic exotic nuclei is of great interest in nuclear physics and astrophysics. In the `southwest' region of $^{78}$Ni, the development of deformation might trigger a major shift in our understanding of explosive nucleosynthesis. To this end, new spectroscopic information on key close-lying nuclei is very valuable. 

\textbf{Purpose:} We intend to measure the isomeric and $\beta$ decay of $^{75}$Co, with one-proton and two-neutron holes relative to $^{78}$Ni, to access new nuclear structure information in $^{75}$Co and its $\beta$-decay daughters $^{75}$Ni and $^{74}$Ni.

\textbf{Methods:} The nucleus $^{75}$Co is produced in relativistic in-flight fission reactions of $^{238}$U at the Radioactive Ion Beam Factory in the RIKEN Nishina Center. Its isomeric and $\beta$ decay are studied exploiting the BigRIPS and EURICA setups.

\textbf{Results:} We obtain partial $\beta$-decay spectra for $^{75}$Ni and $^{74}$Ni, and report a new isomeric transition in $^{75}$Co. The energy [$E_{\gamma}=1914(2)$ keV] and half-life [$t_{1/2}=13(6)$ $\upmu$s] of the delayed $\gamma$ ray lend support for the existence of a $J^{\pi}=(1/2^-)$ isomeric state at 1914(2) keV. A comparison with PFSDG-U shell-model calculations provides a good account for the observed states in $^{75}$Ni, but the first calculated $1/2^-$ level in $^{75}$Co, a prolate $K=1/2$ state, is predicted about 1 MeV below the observed $(1/2^-)$ level.

\textbf{Conclusions:} The spherical-like structure of the lowest-lying excited states in $^{75}$Ni is proved. In the case of $^{75}$Co, the results suggest that the dominance of the spherical configurations over the deformed ones might be stronger than expected below $^{78}$Ni. Further experimental efforts to discern the nature of the $J^{\pi}=(1/2^-)$ isomer are necessary. 

\end{abstract}

\maketitle
%\tableofcontents

\section{Introduction}

The neutron-rich region approaching $^{78}$Ni, with 28 protons and 50 neutrons, is in the spotlight of the most important radioactive-ion beam facilities as this nucleus is the most exotic doubly magic one ever synthesized in the laboratory \cite{Hos05,Xu14,Sum17}. As more access to new structural features is obtained in nearby nuclei \cite{Sod15,Ben15,San15,Oli17,Wel17,Sha17,Sah17,Mor18,Lok20,Gar20}, $^{78}$Ni appears to be the last spherical system prior to the hitherto unattainable domain of the \textit{r}-process reaction path \cite{Now16}. Indeed, the most advanced theoretical calculations presently available \cite{Now16,Ots05} predict the coexistence of spherical and prolate deformed shapes in $^{78}$Ni, with a $0^+$ deformed bandhead lying at about the same excitation energy of the first $2^+$ state, or even below \cite{Now16,Tan19}. Interestingly, first experimental fingerprints for the existence of the deformed configuration have recently been provided by Taniuchi \textit{et al.} \cite{Tan19}, who have proposed a 2.91-MeV deformed $(2^+)$ candidate just above the spherical $(2^+_1)$ state at 2.60 MeV. Just by adding a few neutrons or removing a few protons to the doubly magic system, the prolate-deformed configuration is expected to drop below the spherical one and become yrast. Such an inversion, with $^{78}$Ni as the doorway to the new ground-state deformation region, might have a substantial effect on the theoretical predictions on the location of the neutron drip line, as deformed systems are expected to be more tightly bound \cite{Tsu20}, hence making a difference to our understanding of the \textit{r}-process nucleosynthesis pathways.

The robustness of the $Z=28$ closed shell and the coexistence of deformed and spherical shapes in the neutron-rich $\nu g_{9/2}$ Ni isotopes have been a matter of debate in a number of recent experimental and theoretical works \cite{Per06,Aoi10,Raj12,Tsu14,Mar14,Kol16,Mor16,Mor17,Mor18,Cor18L,Elm19,Ele19,Got20,Bel20,Go20}. Particularly important is the conservation of the seniority quantum number $\upsilon$ \textendash the number of protons or neutrons that are not coupled in pairs to $J=0$ \cite{Talmi,Casten}\textendash~ as it is a good indicator of gap stability. In the Ni nuclei filling the $\nu g_{9/2}$ shell, the seniority is still a good quantum number for a sub-set of solvable eigenstates \cite{Esc06,Qi11,Zam07,Isa08,Qi10,Isa13}, although it is still unknown if the deformation-driving forces might induce mixing of seniorities in close-lying states with equal $J$ \cite{Mor18,Mac17,Qi17}.

The increase in collectivity as protons are removed from the $Z=28$ closed-shell regime has also been deeply investigated \cite{Cra13,Ben15,San15,Lid15,Mor17,Cor19,Gad19,Lok20,Can20}, indicating the development of a new island of inversion around $N=40$ that extends beyond the harmonic-oscillator shell. Theoretically, the development of deformation around $N=40$ appears to be driven not only by the variation in the number of protons and neutrons as one moves away from stability, but by many particle-hole excitations across energy gaps eventually induced by the proton-neutron tensor component of the nuclear force \cite{Ots16,Tog16,Ots19,Ots20}, which causes a reduction of the $\pi f_{7/2} - \pi f_{5/2}$ spin-orbit splitting as neutrons occupy the $\nu g_{9/2}$ orbital \cite{Ots05,Ots13,Tsu14}. 

\begin{figure*}[t]
	%\vspace*{0.1cm}
	%\hspace*{-0.5cm}
	\centering
	\includegraphics[width=1.\textwidth]{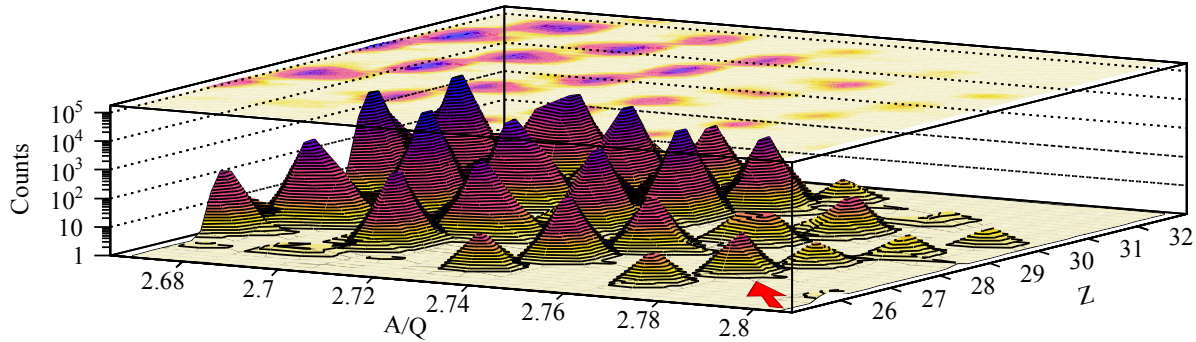} 
	\caption{(Color online) Three-dimensional cluster plot of the nuclei implanted in WAS3ABi as a function of their charge $Z$ and mass-to-charge ratio $A/Q$. The cluster corresponding to $^{75}$Co is indicated by the red arrow.}
	\label{fig1}
\end{figure*}

The convergence of the $N=40$ island of inversion with a newly predicted region of deformation around $N=50$ has been theorized recently \cite{Now16}. This phenomenon is comparable to the merging of the $N=20$ and $N=28$ closed shells, with similar underlying mechanisms driving the onset of deformation and the disappearance of the classical shell closures. Although the observation of a deformed candidate state in coexistence with the normal spherical shape in $^{78}$Ni supports this prediction, more experimental information on lighter $Z \leq 28$ nuclei towards $N=50$ is needed to fully comprehend how the shell structure evolves in the neutron-rich region below $Z=28$, and if there is a new $N=50$ island of inversion in coalescence with the one at $N=40$.

With these goals in mind, the isomeric and $\beta$ decay of $^{75}$Co, with one proton and two neutron holes relative to the $^{78}$Ni doubly magic core, were investigated following the in-flight fission of a relativistic $^{238}$U beam on a thin natural Be target at the Radioactive Isotope Beam Factory (RIBF) at RIKEN, Japan. Despite being pinpointed as one of the $A\sim78$ nuclei with a significant impact on $r$-process estimates \cite{Sha16}, the only experimental information reported hitherto in the literature for $^{75}$Co is limited to the half-life $t_{1/2}$ and an upper limit for the $\beta$-delayed one-neutron emission probability $P_{1n}$ \cite{HosmerHalflives2010,Xu14,Go20}. Here, we provide an additional lower $P_{1n}$ limit, of help to extend the experimental databases used by nuclear astrophysicists. For $^{75}$Ni, four $\gamma$ rays at 232 keV, 893 keV, 950(20) keV and 1100(20) keV have recently been reported \cite{Go20,Got20}. While the first two were observed following $\beta$ decay of $^{75}$Co in an in-flight fragmentation experiment in NSCL \cite{Go20}, the latter two were reported in an intermediate Coulomb excitation experiment carried out at RIKEN \cite{Got20}. Of them, the 232-keV, 950(20)-keV and 1100(20)-keV transitions have tentatively been placed in the level scheme of $^{75}$Ni, decaying directly to the ground state from levels with proposed spins and parities $J^{\pi}=(7/2^+)$, $(13/2^+)$, and $(11/2^+)$, respectively. In the three cases the $J^{\pi}$ arises from the $\nu g_{9/2}^n$, $\upsilon=3$ seniority configuration. In the present work, an extended experimental study with new $\gamma$-ray transitions in both $^{75}$Co and $^{75}$Ni is reported. The new spectroscopic information is compared to state-of-the-art large-scale shell-model calculations performed with the code ANTOINE \cite{Cau05} and the PFSDG-U interaction in the $pf$-$sdg$ valence space \cite{Now16,Now21}. On the basis of the experimental and theoretical results presented here, we argue new spin assignments and discuss the evolution of the spherical and deformed configurations in the `southwest' quadrant of $^{78}$Ni.

\section{Experimental details}

The present data were obtained during the EURICA campaign at the RIBF, operated jointly by the RIKEN Nishina Center and the Center for Nuclear Study of the University of Tokyo. The primary beam of $^{238}$U was delivered by the RIKEN accelerator complex, which consisted of a linear injector (RILAC2) and four ring cyclotrons (RRC-fRC-IRC-SRC). The beam energy was 345 MeV/nucleon, with an intensity of approximately $3 \times 10^{10}$ pps. The nucleus $^{75}$Co and other neutron-rich nuclides close to $^{78}$Ni were produced by in-flight fission \cite{Ohn10} on a 3-mm-thick foil of $^{9}$Be. The secondary beam species of interest were separated in both the first and second stages of the BigRIPS magnetic spectrometer \cite{Fuk13} using dipole magnets. The selected fission fragments were identified through the standard $\Delta E$-$B \rho$-$TOF$ method in the second stage of BigRIPS. Beam-line detectors, as fast plastic scintillators, parallel-plate avalanche counters, and a multisampling ionization chamber, allowed for an event-by-event particle identification of the atomic number ($Z$) and the mass-to-charge ratio ($A/Q$) of the secondary-beam products. 

The radioactive ion beam was conducted through the zero-degree spectrometer (ZDS) \cite{zerodegree} to the EURICA $\beta$-decay spectroscopy station, consisting of the active beam stopper WAS3ABi \cite{Nis12} and the $\gamma$-ray spectrometer EURICA \cite{Sod13}. Since the radioactive nuclei identified in BigRIPS were very energetic, it was necessary to place a homogeneous aluminium degrader of variable thickness before WAS3ABi in order to adjust the range of the ions of interest within the implantation device.

The silicon array WAS3ABi was not only used to stop the radioactive nuclei but also to detect electrons and other charged particles emitted in their decay. WAS3ABi consisted of eight layers of 1-mm-thick double-sided silicon strips detectors (DSSSD) with an interspace of 0.5 mm between them. Each DSSSD had an active area of $60\times40$ mm$^2$ and was segmented into 60 vertical strips and 40 horizontal strips, providing a total of 2400 pixels of 1-mm pitch each. The WAS3ABi DAQ system recorded the pixel position, time, and energy information of the implanted fission products and the emitted $\beta$ electrons. Standard analog electronics were used to read the energy and time signals of each strip, optimized for the energy range of $\beta$ particles. Meanwhile, in-flight fission fragments released around 1 GeV in the detector, overflowing the energy signals of the implantation strip and the neighboring ones. The position of implantation $(X,Y)$ was then defined by the horizontal and vertical strips with the fastest time signal \cite{Nis13}.

The EURICA array was set up surrounding the active beam stopper and was used to record the energy and time of $\gamma$ rays during a time window of up to 110 $\upmu$s after the detection of an implantation or $\beta$ electron. As a result, the setup was sensitive to isomeric lifetimes ranging from several ns to several hundred $\upmu$s. EURICA was made of 84 high-purity germanium (HPGe) crystals, arranged in 12 clusters of seven crystals packed closely at an average distance of 22 cm from WAS3ABi. An absolute detection efficiency of approximately 11\% at 662 keV was achieved after applying a standard add-back routine \cite{Mor16}.

\section{Analysis procedure}
\label{sec:anproc}

In the off-line analysis, implantation-like events were defined by an overflow energy signal in at least one horizontal and one vertical strip of WASA3Bi. These were requested to be in coincidence with a high-energy signal in the last fast plastic scintillator of ZDS and in anticoincidence with any signal above threshold registered in a $\beta$ detector placed behind WAS3ABi \cite{Orr20}. In this way, secondary reaction products generated during the implantation process were rejected to a large extent. The DSSSD of implantation was then identified as the last one in which a horizontal and a vertical strip were overflowed. On the other hand, electron-like events were defined by non-overflow energy signals above $\beta$ threshold ($\sim$ 50 keV) in anticoincidence with the last plastic scintillator of ZDS. Since a $\beta$ electron typically fired several strips before leaving WAS3ABi, the total energy released in each DSSSD was obtained from the sum of the energies of adjacent strips within a 8-$\upmu$s time gate. The $(x,y)$ position of the $\beta$-like particles was then computed as the energy-weighted average of the fired horizontal and vertical strips.

Once defined, implantations and $\beta$ particles were correlated in position and time. In the present analysis, the spatial correlations were restricted to the DSSSD of implantation, and the maximum transverse distance was fixed to $\rho= \sqrt{(x-X)^2 + (y-Y)^2} = \sqrt2$ pixels. The time window was set to $t=135$ ms, corresponding to about five half-life periods of $^{75}$Co \cite{Xu14}. Additional prompt-time correlations with $\gamma$ rays were defined to explore the structure of implanted and descendant nuclei. These were set according to the expected nuclear half-lives, and had maximal time windows of 800 ns for non-isomeric $\beta$-delayed transitions and $\sim$ 50 $\upmu$s for isomeric transitions. For the study of coincident $\gamma$ rays, a 400-ns time window was set. Background contributions from randomly correlated events were evaluated by applying the so-called backward-time technique \cite{Mor13}, which exploits correlations between implantations and preceding $\beta$ electrons using the same conditions as for the normal correlations.

\begin{figure*}
	%\vspace*{0.1cm}
	%\hspace*{-0.5cm}
	\centering
	\includegraphics[width=\textwidth]{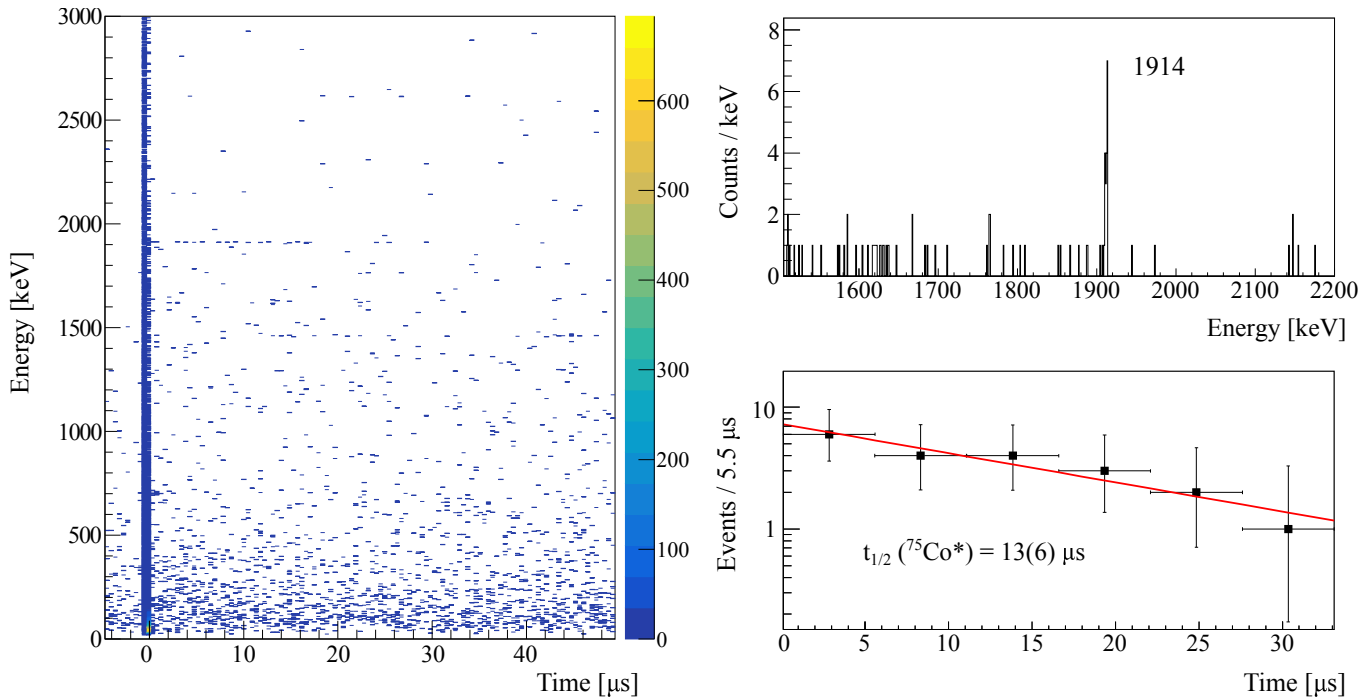} 
	\caption{(Color online) Left: Energy-time matrix showing the $\gamma$ rays detected after implantation of $^{75}$Co. The $\gamma$-ray energy is plotted against the ion-$\gamma$ time difference. Top right: Projection of the matrix on the energy axis for a time window of $\sim$35 $\upmu$s. Bottom right: Projection of the matrix on the time axis, gated on the 1914-keV transition. The fit to a single exponential function is shown in red. The resulting half-life is indicated. See text for details.}
	\label{fig2}
\end{figure*}

\section{Results}

The distribution of implantations as a function of $Z$ and $A/Q$ is shown in the three-dimensional plot of Fig. \ref{fig1}, where the red arrow pointing to the $^{75}$Co nuclei illustrates the good quality of the particle identification. In total, $\sim1.8\times10^4$ ions of $^{75}$Co were implanted in WAS3ABi.

\subsection{Isomeric spectroscopy}

The two-dimensional energy-time matrix of the $\gamma$ rays detected within approximately $50$ $\upmu$s after the detection of $^{75}$Co implantation events is shown in the left panel of Fig. \ref{fig2}. A newly identified isomeric $\gamma$ transition can be seen at about 1900 keV. Matrix projections on the Y (energy) and X (time) axis are shown in the right top and bottom panels of the figure, respectively. The resulting energy for the isomeric transition is $E_{\gamma}=1914(2)$ keV. Due to the scarce statistics, the half-life has been obtained from an unbinned maximum likelihood fit to a single exponential time distribution function describing the decay time behavior of the $\gamma$ events registered after the so-called \textit{prompt flash} (which is visible at time $\sim 0$). No background contributions have been considered in the fitting procedure due to the absence of random $\gamma$ events in the region of interest of the energy-time matrix. The resulting half-life is $t_{1/2}(^{75}$Co$^*)=13(6)$ $\upmu$s. The reported uncertainty is only statistical and has been evaluated using the RooFit package \cite{Ver03} with the MINOS method for determination of error parameters \cite{Minuit}.

\begin{table*}[]
	\centering 
	\renewcommand{\arraystretch}{1.2}
	\renewcommand{\tabcolsep}{0.7 cm}
	\footnotesize
	\caption{List of $\gamma$ transitions observed in the $\beta$ decay of $^{75}$Co and placed in the level scheme of any of the descendant nuclei. The nuclei to which the transitions are assigned, the $\gamma$-ray energies, the excitation energies of initial and final states, and the absolute $\gamma$-ray intensities are given. As well, the $\beta$($\gamma\gamma$) coincidence relations are indicated for each transition. An asterisk highlights those cases where the coincidence relation is established by only one observed count, but the coincidences with other transitions of the mutual $\gamma$ cascade are observed.} 
	\label{table1}
\begin{tabular}{l l l l l l}
\hline
\hline
Nucleus   & E$_{\gamma}$ (keV) & E$_x^i$ (keV) & E$_x^f$ (keV) & I$_{\gamma}$ (\%) & Coincident $\gamma$ rays    \\
\hline
$^{75}$Ni & 231.8(9)                  & 231.8(9)                          & 0                                 & 41(3)                    & 740.8, 891.4, 1632.2        \\
$^{75}$Ni & 416.6(12)                 & 1461.1(17)                        & 1044.6(11)                        & 5.8(9)                  & 1044.6, 1596                \\
$^{75}$Ni & 740.8(10)                 & 972.7(9)                         & 231.8(9)                          & 7.1(11)                  & 231.8, 891.4                \\
$^{75}$Ni & 866.5(11)                 & 1911.1(16)                        & 1044.6(11)                        & 2.6(7)                   & 1044.6                      \\
$^{75}$Ni & 891.4(10)                 & 1864.4(8)                         & 972.7(9)                         & 12.5(16)                 & 231.8, 740.8, 972.9         \\
$^{75}$Ni & 972.9(12)                 & 972.7(9)                         & 0                                 & 7.0(12)                  & 891.4                       \\
$^{75}$Ni & 1044.6(11)                & 1044.6(11)                        & 0                                 & 8.2(13)                  & 416.6, 866.5, 1596*          \\
$^{75}$Ni & 1596(3)                   & 3057(3)                           & 1461.1(17)                        & 4.0(10)                  & 416.6, 1044.6*               \\
$^{75}$Ni & 1632.2(13)                & 1864.4(8)                         & 231.8(9)                          & 24(3)                    & 231.8                       \\
$^{75}$Ni & 1865.1(10)                & 1864.4(8)                         & 0                                 & 1.6(7)                   & \---                           \\
$^{74}$Ni & 226.0(5)                  & 2606(2)                           & 2380(2)                           & 1.5(4)                   & 616.7, 738.3, 1024.6*        \\
$^{74}$Ni & 616.7(13)                 & 2380(2)                           & 1762.9(16)                        & 5.8(10)                  & 226.0, 738.3, 1024.6        \\
$^{74}$Ni & 738.3(10)                 & 1762.9(16)                        & 1024.6(12)                        & 7.9(12)                  & 226.0, 616.7, 1024.6        \\
$^{74}$Ni & 1024.6(12)                & 1024.6(12)                        & 0                                 & 17(2)                    & 226.0*, 616.7, 738.3, 1079.7 \\
$^{74}$Ni & 1079.7(10)                & 2104.2(16)                        & 1024.6(12)                        & 3.2(8)                   & 1024.6                      \\
\hline
\hline
\end{tabular}
\end{table*}

\subsection{$\beta$-decay spectroscopy}

\begin{figure*}[t]
	%\vspace*{0.2cm}
	%\hspace*{-0.5cm}
	\centering
	\includegraphics[width=0.8\textwidth]{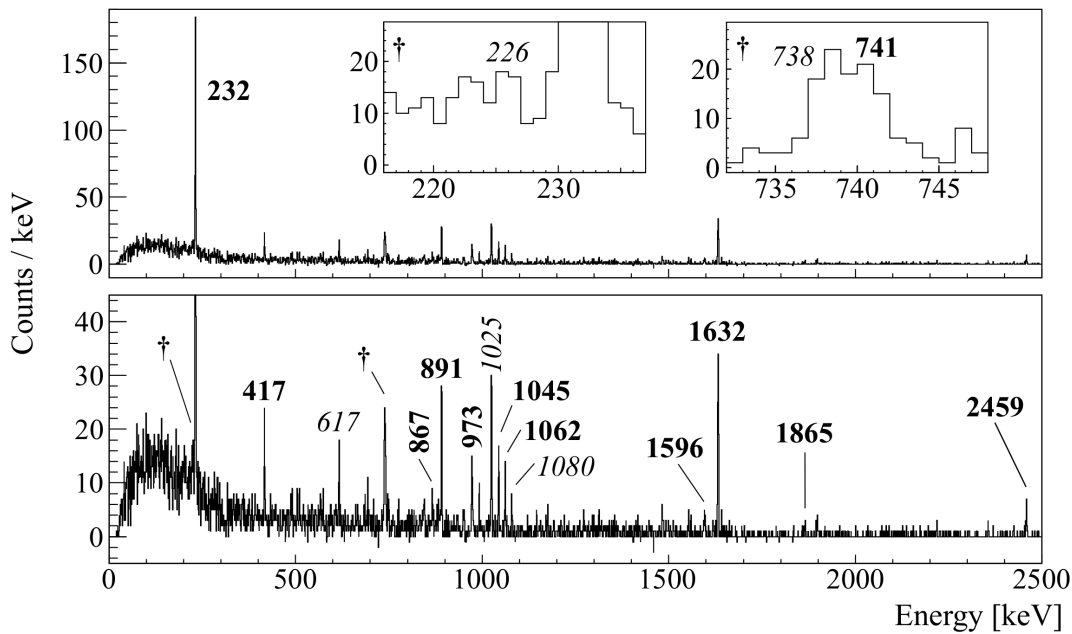}
	\caption{The $\beta$-delayed $\gamma$-ray energy spectrum following implantations of $^{75}$Co nuclei during a time interval of 135 ms. The panels present two different ranges of the y axis to facilitate the observation of weak $\gamma$ rays. The transitions assigned to the $\beta$ ($^{75}$Ni) and $\beta_n$ ($^{74}$Ni) daughters are marked in bold and italics, respectively. Expanded inset spectra are shown for the $\gamma$ rays marked with a dagger.}
	\label{fig3}
\end{figure*}

\begin{figure}[h]
    \centering
	\includegraphics[width=0.48\textwidth]{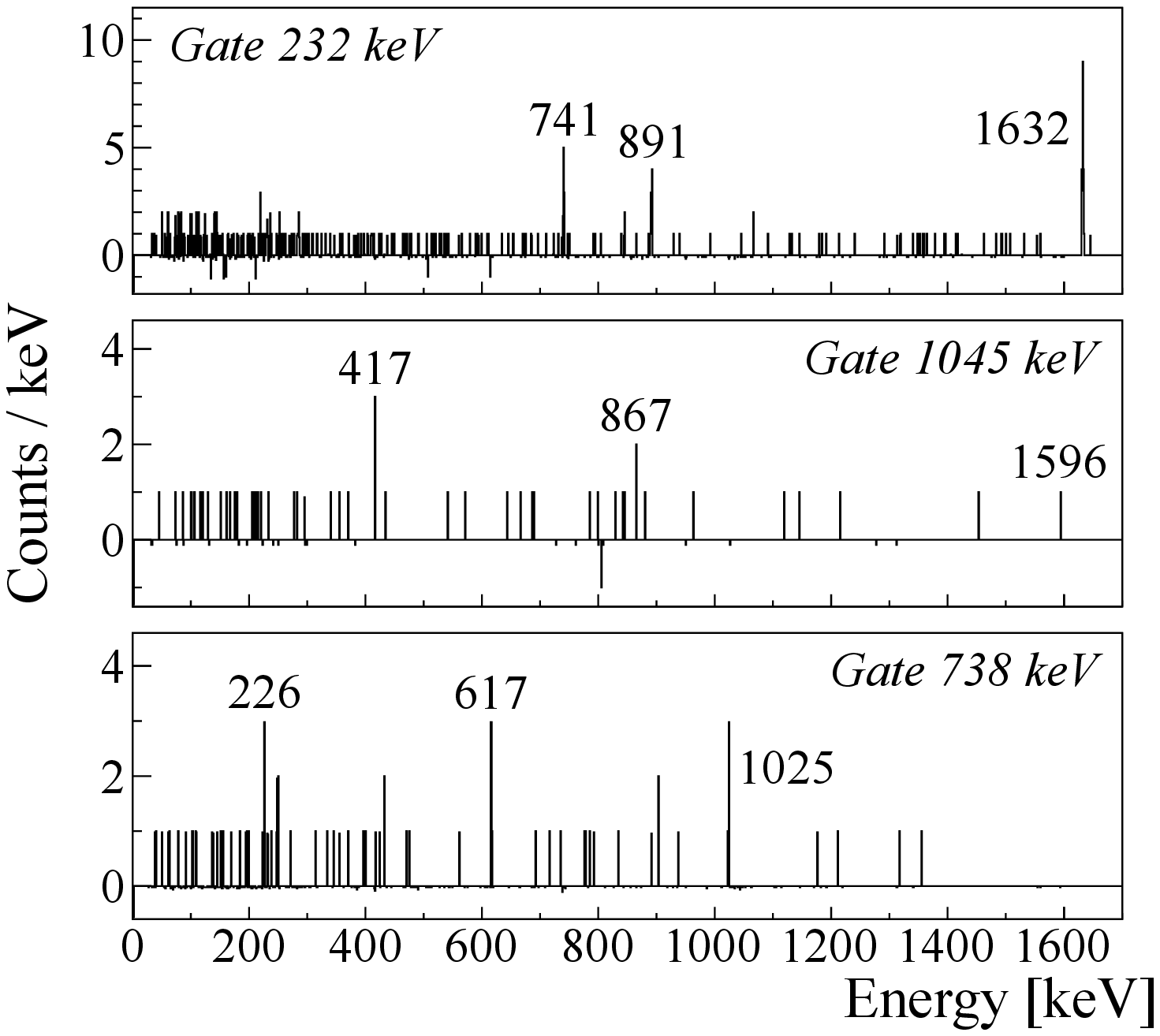}
	\caption{From top to bottom, $\beta(\gamma\gamma)$ coincidence spectra gated on the 232-, 1045-, and 738-keV $\gamma$ transitions.}
	\label{fig4}
\end{figure}

The singles $\beta$-delayed $\gamma$-ray energy spectrum resulting from the analysis procedure described in Sec. \ref{sec:anproc} is shown in Fig. \ref{fig3}. There, the most intense transitions attributed to the $\beta$ ($^{75}$Ni) and $\beta_n$ ($^{74}$Ni) descendants are labeled in bold and italics, respectively. 
As an example, the coincidence spectra gated on the 232-, 1045-, and 738-keV $\gamma$ transitions are shown in the three panels of Fig. \ref{fig4}. While the first two are attributed to $^{75}$Ni, the third one is assigned to the $\beta_n$ daughter $^{74}$Ni. In all cases, the background contributions have been evaluated and subtracted as described in Ref. \cite{Mor17b}. 

The full list of $\gamma$-ray transitions, absolute $\gamma$ intensities, and $\gamma$-$\gamma$ coincidence relations observed following $\beta$ decay of $^{75}$Co is provided in Tables \ref{table1} and \ref{table2}. While Table \ref{table1} shows transitions placed in the level scheme of any of the daughter nuclei ($^{74}$Ni and $^{75}$Ni), Table \ref{table2} shows the list of $\gamma$ rays attributed to $^{75}$Ni that have not been placed in the level scheme due to the absence of coincident transitions. It is to note here the large intensity of the $\gamma$ rays at 1061.8(10) keV and 2458.8(15) keV.

\begin{table}[h!]
	\centering 
	\renewcommand{\arraystretch}{1.2}
	\renewcommand{\tabcolsep}{0.7 cm}
	\footnotesize
	\caption{Transitions attributed to $^{75}$Ni that have not been placed in the level scheme. The $\gamma$-ray energies and the absolute $\gamma$-ray intensities are shown.}
	\label{table2}
\begin{tabular}{l l l}
\hline
\hline
Nucleus              & E$_{\gamma}$ (keV) & I$_{\gamma}$ (\%) \\
\hline
$^{75}$Ni   & 491.3(9)              & 1.5(5)                \\
$^{75}$Ni   & 566.8(11)              & 1.5(5)                \\
$^{75}$Ni   & 686.1(5)               & 1.4(5)                \\
$^{75}$Ni   & 1061.8(10)             & 5.8(11)               \\
$^{75}$Ni   & 1145.8(9)             & 1.8(6)                \\
$^{75}$Ni   & 1175.9(12)             & 2.2(7)                \\
$^{75}$Ni  & 1313.1(7)              & 1.4(6)                \\
$^{75}$Ni  & 1559.2(8)              & 1.5(6)                \\
$^{75}$Ni  & 2219.0(10)             & 2.0(8)                \\
$^{75}$Ni  & 2355.7(12)             & 1.6(7)                \\
$^{75}$Ni  & 2458.8(15)             & 5.6(14)              \\
\hline
\hline
\end{tabular}
\end{table}

The partial level schemes corresponding to the internal and $\beta$ decay of $^{75}$Co are shown in Fig. \ref{fig5}. 
The ordering of the transitions following $\beta$ decay is proposed according to $\gamma$-ray intensity balances, $\gamma$-$\gamma$ prompt coincidence relations, and $\gamma$-ray energy sum matchings according to the information displayed in Table \ref{table1}. Apparent $\beta$ feedings and/or log$ft$ values are indicated at the left of the level schemes. These can be considered as upper and lower limits, respectively, due to the large pandemonium effect \cite{Har77} expected in odd-mass nuclei with large $Q_{\beta}$ values, as is the case of $^{75}$Co, with $Q_{\beta}=14380(580)$ keV \cite{Wan17}. In the figure, arrow widths are proportional to transition intensities, and tentative spins and parities are shown in parentheses on the left of each level.

In the $\beta$-decay daughter $^{75}$Ni, the levels at 1864.4(8) keV, 972.7(9) keV and 231.8(9) keV are established based on the observation of three independent $\gamma$-ray cascades to the ground state, namely the 891-973-, 891-741-232- and 1632-232-keV $\gamma$ cascades, and the direct ground-state transition at 1865 keV. The location of the level at 1044.6(11) keV is based on two arguments: first, the 1045-keV transition is the most intense one in the set of coincident $\gamma$ rays formed by the 1045-, 417-, 867-, and 1596-keV peaks (see Table \ref{table1}). And second, its energy could match with the tentative direct ground-state transition at 1100(20) keV reported by Ref. \cite{Got20}. The placement of the states at 1461.1(17) keV, 1911.1(16) keV, and 3057(3) keV, built up on top of the 1044.6(11)-keV level, is proposed according to the coincidence relationships indicated in Table \ref{table1}. Meanwhile, the observed states at 1024.6(12), 1762.9(16), 2104.2(16), 2380(2), and 2606(2) keV in the $\beta_n$ daughter $^{74}$Ni were previously reported in the direct decay of $^{74}$Co to $^{74}$Ni \cite{Mor18} and identified as following the $(8^+_1) \rightarrow (6^+_1) \rightarrow (4^+_1) \rightarrow (2^+_1) \rightarrow 0^+$ and $(4^+_2) \rightarrow (2^+_1) \rightarrow 0^+$ $\gamma$ cascades connecting states of seniorities $\upsilon=2$ and $\upsilon=4$. A lower limit for the $\beta$-delayed one-neutron emission probability of $^{75}$Co has been deduced from the absolute intensity of the $(2^+_1)\rightarrow 0^+$ transition at 1025 keV, resulting in $P_{1n}\geq15\%$. This value is in good agreement with the upper limit reported in the literature, $P_{1n}\leq16\%$ \cite{HosmerHalflives2010}, and points to a rather low ground-state feeding in the $\beta_n$ decay $^{75}$Co $\rightarrow$ $^{74}$Ni. 

\section{Discussion}

\subsection{$\beta$ decay of $^{75}$Co to $^{75}$Ni}

The ground state of the odd-even parent nucleus $^{75}$Co is proposed to have a tentative $J^{\pi}=(7/2^-)$ based on the lowest-lying $\pi f_{7/2}^{-1}$ proton-hole configuration. This assignment is in accordance with the tentative $J^\pi=(7/2^-)$ attributed to the ground states of the lighter odd-even isotopes $^{71}$Co and $^{73}$Co \cite{Raj12,Lok20}. Meanwhile, the main contribution to the ground-state wave function in the daughter nucleus $^{75}$Ni is expected from the unpaired neutron in the $\nu g_{9/2}$ shell, resulting in $J^\pi = (9/2^+)$. The assignment is supported by the systematics of lighter neutron-rich $\nu g_{9/2}$ even-odd Ni isotopes \cite{Mol99,Raj12} and the recent experimental studies of $^{75}$Ni \cite{Go20,Got20}.

The strong $\beta$ feeding to the excited state at 1864.4(8) keV, $I_{\beta}=38(3)\%$, and the corresponding log\textit{ft} = 4.7(1), provide a robust proof for the occurrence of an allowed Gamow-Teller (GT) decay from the $\pi f_{7/2}^{-1}$ ground state of $^{75}$Co. As the most energetic single-particle GT transition occurring in the region of $^{78}$Ni transforms a neutron in the $\nu f_{5/2}$ orbital into a proton in the $\pi f_{7/2}$ shell, the associated wave function in the final 1864.4(8)-keV state of $^{75}$Ni is expected to have a large $\nu f_{5/2}^{-1}$ contribution, resulting in a tentative spin and parity $J^\pi = (5/2^-)$. Further support for the (5/2$^-$) assignment to this level comes from the systematic comparison with the $\beta$-decay level schemes of lighter even-odd Ni isotopes (see Refs. \cite{Raj12,Mol99}), which shows that the most probable $\beta$-decay transition populates the yrast $(5/2^-)$ level. For the first excited state at 231.8(9) keV in $^{75}$Ni, we propose $J^\pi=(7/2^+)$ despite an observed (apparent) feeding of $I_{\beta}<14\%$. Our assignment is in agreement with the work of Go \textit{et al.} \cite{Go20} and it is equally based on the similarity with the excitation energies of the $(7/2^+_1)$ states in $^{71}$Ni (281 keV) and $^{73}$Ni (239 keV) \cite{Raj12}. This spin and parity arises mainly from the coupling of the first $(2^+)$ state in the $^{74}$Ni core to the unpaired $\nu$g$_{9/2}$ neutron, and has as main configuration $\nu g_{9/2}^{n}$. Apart from this level, additional states with $J^{\pi}=3/2^+$, $5/2^+$, $9/2^+$, $11/2^+$, and $13/2^+$ are expected to arise from the $\nu g_{9/2}^n$ multiplet. Excepting the $9/2^+_2$ level, the rest of states are expected to lie at excitation energies between 1 MeV and 1.5 MeV (see Refs. \cite{Raj12,Got20} and discussion in Sec. \ref{sec:SM}).

The almost nonexisting $\beta$ feedings to the levels at 972.7(9) keV and 1044.6(11) keV indicate that, more likely, they are not fed by allowed GT transitions but through internal $\gamma$ feeding, in accordance with the forbiddenness of a $\nu g_{9/2}$ $\rightarrow$ $\pi f_{7/2}$ single-particle $\beta$ transition. In the case of the 972.7(9)-keV level, the energy matches well with that of the reported $(13/2^+)$ state at 950(20) keV \cite{Got20}. The observation of a prompt, strong $\gamma$ ray at 891 keV connecting the $(5/2^-)$ level with this state, though, rules out spin assignments higher than $9/2$. Hence, the only positive-parity states remaining at this excitation energy are $J^{\pi}=3/2^+$ and $5/2^+$. In Table \ref{table3}, the quotients $R$ between the single-particle branching ratios of the $\gamma$ rays decaying from the 972.7(9)-keV level are shown for each possible $J^{\pi}$ and transition multipolarities. The $\gamma$-ray energies are 740.8(10) keV and 972.9(12) keV. A comparison with the experimental quotient of these two transitions, $R=1.0(4)$, provides support for a $J^\pi = (5/2^+)$ assignment. 
Based on this, and given the previous spectroscopic information from relativistic Coulomb excitation (see Fig. 2 of Ref. \cite{Got20}), we propose the 1044.6(11)-keV state to have $J^{\pi}=(11/2^+)$ or $(13/2^+)$. 

\begin{figure*}
	%\vspace*{0.1cm}
	%\hspace*{-0.3cm}
	\centering
	\includegraphics[width=0.95\textwidth]{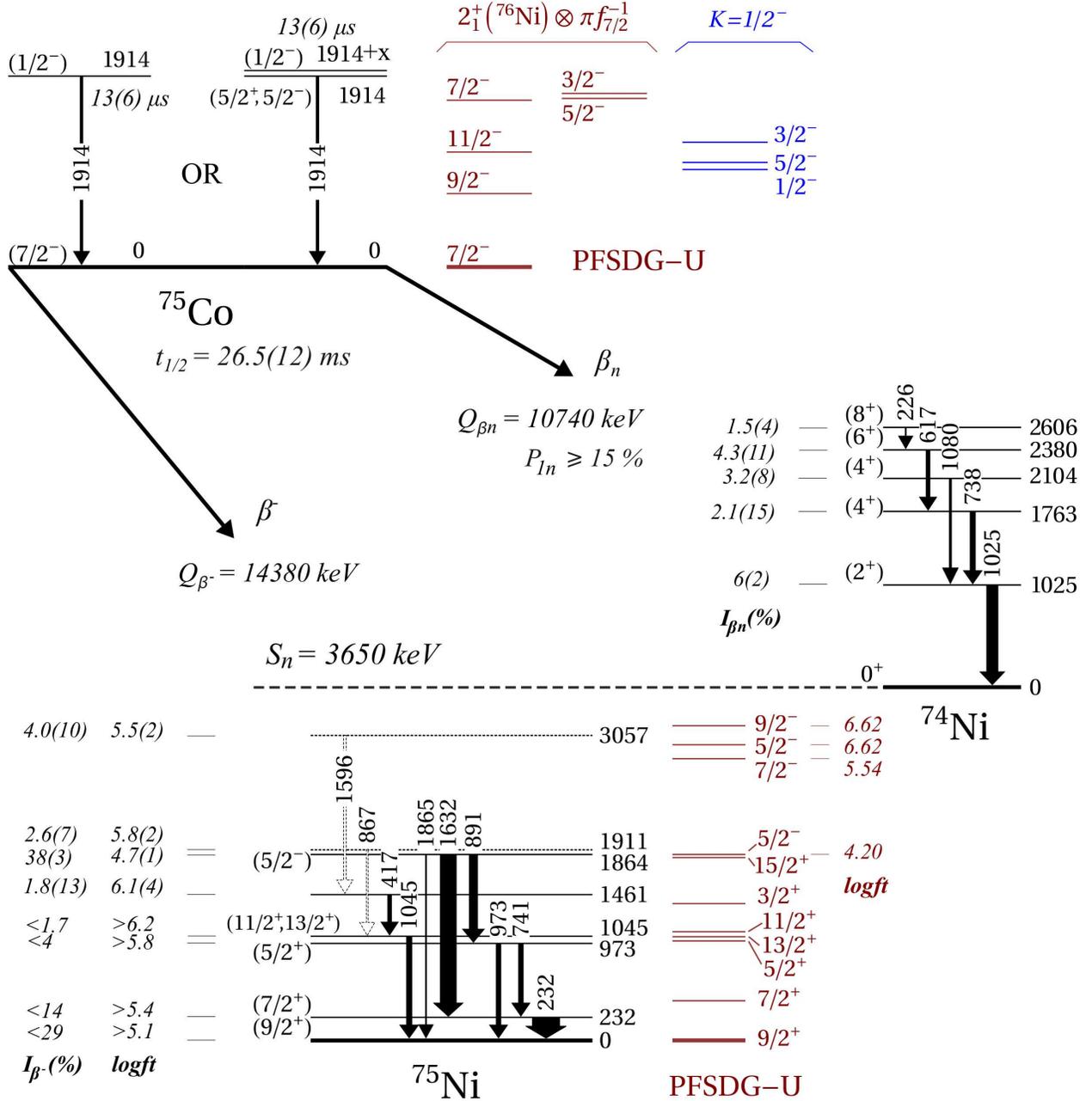} 
	\caption{(Color online) Experimental isomeric and $\beta$-decay level schemes of $^{75}$Co extracted from the present work. The energies of the levels are given in keV. The thickness of the arrows is proportional to the intensities of the transitions connecting the states. Spins and parities, apparent $\beta$ feedings and log$ft$ values are indicated at the sides of the levels. Theoretical states, obtained with the PFSDG-U interaction \cite{Now16,Now21}, are indicated in red and blue. The half-life of $^{75}$Co is reported in Ref. \cite{Xu14} while the $Q_{\beta}$, $Q_{\beta n}$, and $S_n$ values are taken from Ref. \cite{Wan17}. See text for details.}
	\label{fig5}
\end{figure*}

The strong 1061.8(10)-keV transition observed in the singles $\gamma$ spectrum of Fig. \ref{fig3} could very likely de-excite the remaining $J^{\pi}=(13/2^+)$ or $(11/2^+)$ level to the ground state, as its energy matches well with those of the calculated $11/2^+$ and $13/2^+$ levels (see Fig. \ref{fig5}). However, such high positive-parity spin states cannot be directly fed from $\beta$ decay of the $^{75}$Co ground state and no coincident $\gamma$ rays have been observed for the 1061.8(10)-keV transition. The 1061.8(10)-keV $\gamma$ ray could alternatively connect the $(5/2^-_1)$ state with the missing $1/2^-_1$ level, which is expected to have $\nu p_{1/2}^{-1}$ as main configuration and is predicted about 700 keV below by the LNPS \cite{Len10} and the calculations of Ref. \cite{Raj12}. The nonobservation of coincident $\gamma$ rays with the 1061.8(10)-keV transition could then be ascribed to the possible $\beta$-decaying character for the final $(1/2^-)$ level. It is to note, though, that the corresponding activity has not been identified in the $\beta$-decay half-life study carried out from ion-$\beta$ time correlations using the present data. Similarly, we have not found evidence for the internal $\gamma$ decay of a $(1/2^-)$ candidate, maybe due to the limited statistics available or maybe because the level lies at a higher excitation energy, as discussed for its lighter neighbor $^{73}$Ni \cite{Raj12}. As a consequence, the 1061.8(10)-keV transition has not been placed in the level scheme shown in Fig. \ref{fig5}.

\begin{table}
	\centering 
	\renewcommand{\arraystretch}{1.2}
	\renewcommand{\tabcolsep}{0.2 cm}
	\footnotesize
	\caption{Possible spins and parities $J^{\pi}$ for the state at 972.7(9) keV, multiplicities of the de-exciting transitions at 740.8(10) keV and 972.9(12) keV, and corresponding fractions of branching ratios $R$ derived from single-particle estimates. See text for details. } 
	\label{table3}
	\begin{tabular}{ l l l l}
		
		\hline
		\hline
		
		$J^{\pi}$ & Mult & Mult & \multirow{2}{*}{ R$\left[ \frac{ BR_{\gamma}^{973} }{ BR_{\gamma}^{741} } \right]  $ }  \\
		
		[E$_x$(973 keV)] & [E$_{\gamma}$(741 keV)] & [E$_{\gamma}$(973 keV)] & \\
		\hline
		5/2$^+$ &	$M1$ &	$E2$	& $1.60\times10^{-3}$ \\
		%\hline
		5/2$^+$ &	$E2$ &    $E2$	& $3.90$ \\
		%\hline
		3/2$^+$ &	$E2$ &	$M3$	& $5.00\times10^{-7}$ \\

		\hline
		\hline
		
	\end{tabular}
	
\end{table}

\subsection{Isomeric decay of $^{75}$Co}
\label{sec:isomer}

The absence of transitions in coincidence with the delayed 1914(2)-keV $\gamma$ ray in $^{75}$Co (see Fig. \ref{fig2}) leads to two possible scenarios. In the first one, the observed transition directly connects an isomeric level at 1914(2) keV with the $J^{\pi}=(7/2^-)$ ground state. In such a case, the measured $\gamma$-ray lifetime, $t^{\gamma}_{1/2}=13(6)$ $\upmu$s, suggests an $M3$ or $E4$ character. An $E4$ nature can be rejected on the basis that the spin and parity of the initial level then would have to be $J^{\pi}=15/2^-$, and a faster decay path would be opened through the $11/2^-$ state that is expected at about 1 MeV from systematics of the lighter $\nu g_{9/2}$ odd-even Co isotopes \cite{Lok20} and the present PFSDG-U calculations (see Fig. \ref{fig5}). For an $M3$ character, $J^{\pi}=13/2^-$ or $1/2^-$ are possible. Of these, $J^{\pi}=13/2^-$ would as well find a faster decay path through the $11/2^-$ or $9/2^-$ levels. Hence, only a $J^{\pi}=1/2^-$ state could result in an isomeric decay to the $J^{\pi}=(7/2^-)$ ground state. This is the first option presented for the experimental level scheme of $^{75}$Co in Fig. \ref{fig5}.

In the second scenario, the delayed transition may remain unobserved if it is of low energy and has a high conversion coefficient. With the current setup, this is more likely to happen for $E2$ or $M2$ transitions with energies below 50-60 keV \cite{Naw12,Got12}. Then, the observed 1914(2)-keV $\gamma$ ray would follow in the subsequent decay to the ground state. Taking a look at the lowest-lying states expected by the PFSDG-U calculations, a possible decay sequence would be $(1/2^-) \rightarrow (5/2^{\pm}) \rightarrow (7/2^-_{g.s.})$. This option is also indicated in Fig. \ref{fig5}. In this latter case, given the large energy difference of the $E2$/$M2$ and $M3$ transitions, one would expect to detect as well a competing $M3$ branch to the ground state. Therefore, the non-observation of two close-lying $\gamma$ rays in Fig. \ref{fig2} lends support for the first interpretation, i.e., that the 1914(2)-keV transition more likely connects the first $(1/2^-)$ level with the $(7/2^-)$ ground state.

\subsection{Comparison with shell-model calculations}
\label{sec:SM}

The theoretical level schemes shown in Fig. \ref{fig5} for $^{75}$Co and $^{75}$Ni have been obtained with shell-model (SM) calculations using the PFSDG-U interaction in a valence space consisting of the full $pf$ shell for protons and the full $pf-sdg$ shell for neutrons \cite{Now16,Now21}. In general, we find a good agreement between the observed yrast levels in $^{75}$Ni and their calculated counterparts, with an accuracy below 200 keV. Regarding the 972.7(9)-keV level, the calculations also support a $(5/2^+_1)$ assignment from comparison of the reduced transition strengths, with $B(E2;5/2^+_1\rightarrow 9/2^+_1)\approx48$ $e^2fm^4$, $B(E2;5/2^+_1\rightarrow 7/2^+_1)\approx1.4$ $e^2fm^4$, and $B(M1;5/2^+_1\rightarrow 7/2^+_1)\approx0.007$ $\mu_{N}^2$. In the present calculations, the neutron and proton effective charges used for the electric quadrupole operator $E2$ are $\varepsilon_n=0.46$ and $\varepsilon_p=1.31$ \cite{Now16,Now21}. With them, the $B(E2)$ strengths predicted for the $11/2^+$ and $13/2^+$ levels are $B(E2;11/2^+_1\rightarrow 9/2^+_1)\approx42$ $e^2fm^4$ and $B(E2;13/2^+_1\rightarrow 9/2^+_1)\approx56$ $e^2fm^4$, respectively. These theoretical results are of the same order than the calculations named as SM1 and SM2 in Ref. \cite{Got20}, and are similarly systematically lower than the recently measured $B(E2)$ values in $^{75}$Ni \cite{Got20}. The origin of the discrepancies might be due to the population of the $(5/2^+)$ state in the intermediate Coulomb excitation experiment, as it lies close in energy to the $(13/2^+)$ and $(11/2^+)$ candidates proposed by Ref. \cite{Got20}.

\begin{figure*}
	%\vspace*{0.1cm}
	%\hspace*{-0.3cm}
	\centering
	\includegraphics[width=0.92\textwidth]{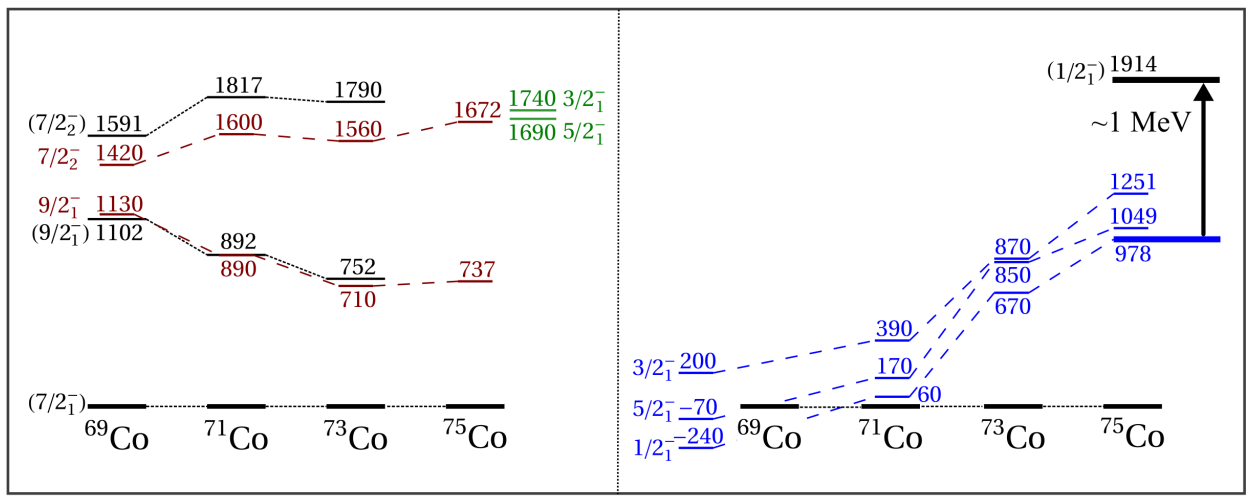}
	\caption{(Color online) Evolution of the experimental (black) and theoretical (red, blue and green) levels discussed here for the odd-mass Co isotopes occupying the $\nu g_{9/2}$ shell. The experimental states are taken from \cite{Lok20} and the present work and the theoretical ones are calculated with the PFSDG-U interaction \cite{Now16,Now21}. The left panel shows states with a spherical-like structure and the right one shows the lowest-lying levels of the $K=1/2$ deformed band.}
	\label{fig6}
\end{figure*}

In addition, we have computed theoretical $B(GT)$ strengths for the $^{75}$Co $\rightarrow$ $^{75}$Ni decay with the PFSDG-U SM calculations. The corresponding log$ft$ values for the lowest allowed states are given in Fig. \ref{fig5}. These have been obtained assuming a standard quenching factor of 0.75. It is to note that the valence space employed is only suited to calculate absolute energies for levels of natural (positive) parity; hence, the negative-parity states derived from the $B(GT)$ computation have been placed by assuming that the excitation energy of the lowest-lying theoretical $5/2^-$ state lies at the excitation energy of the experimental $(5/2^-)$ level, 1864 keV. This assumption is supported by the excellent agreement with the SM prediction discussed in Ref. \cite{Raj12}, which places a $5/2^-$ level with a strong $\sim 76\%$ contribution from the $\nu f_{5/2}^{-1}$ configuration at 1821 keV. 

The calculations clearly indicate an abundance of population to the lowest-lying $5/2^-$ state, with $B(GT)=0.240$ and log$ft=4.20$, in good agreement with the experimentally measured log$ft=4.7(1)$. Theoretically, three negative-parity states are fed by $\beta$ decay at excitation energies around that of the experimental 3057-keV level. Their energies and $J^{\pi}$ values are 2.82 MeV and $7/2^-$, 2.96 MeV and $5/2^-$, and 3.15 MeV and $9/2^-$. The most strongly populated one is the $7/2^-$ level, with $B(GT)= 0.011$ and log$ft=5.54$. This result is close to the experimental log$ft$ value of the 3057-keV state, log$ft= 5.5(2)$. However, as the location of this level is only tentative due to the scarcity of $\beta(\gamma\gamma)$ coincidences, no spin-parity assignment is proposed. Another argument is that, given the large pandemonium effect influencing our data, none of the other two spins and parities can be discarded, even if the calculated log$ft$ values increase up to log$ft= 6.62$. 

Note that the 2458.8(15)-keV transition could connect any of the above states with the 232-keV, $(7/2^-)$ level through a strong $M1$ decay; however, based on the measured $\gamma$ efficiencies and intensities, one would expect to observe between three and four counts at 232 keV in the $\beta(\gamma\gamma)$ spectrum of the 2459-keV peak that do not appear. On the other hand, the $7/2^-_2$ and $5/2^-_2$ states could as well feed the $(5/2^+)$ level at 973 keV through a strong 2459-keV $M1$ transition. In such a case the non-observation of coincidences with the transitions de-exciting the 973-keV level could be explained in terms of $\gamma$ intensity and efficiency arguments, as one would expect one coincident count or less at 741 and 973 keV in the $\beta(\gamma\gamma)$ spectrum of the 2459-keV peak. In this scenario the initial level would lie at about 3432 keV, a higher excitation energy than those predicted by the present SM calculations for the $7/2^-_2$ and $5/2^-_2$ states. Note also that the missing feeding of the 973-keV level, $I_{\beta} < 4\%$, would be slightly smaller than, though still compatible with the observed intensity of the 2459-keV transition, $I_{\gamma} = 5.6(14)\%$. Again, the non-observation of coincidences prevents us from placing the 2458.8(15)-keV $\gamma$ ray in the level scheme of Fig. \ref{fig5}.   

In the case of $^{75}$Co, the PFSDG-U levels are an extension of the SM calculations reported in Ref. \cite{Lok20} for $^{69}$Co, $^{71}$Co and $^{73}$Co using the LNPS \cite{Len10} and PFSDG-U \cite{Now16} interactions. Similar to its lighter neighbours, two well defined structures associated to spherical (red and green) and deformed (blue) shapes are distinguished at low excitation energies in Fig. \ref{fig6}. The first is related to the coupling of the $\pi f_{7/2}^{-1}$ proton hole to the first $2^+$ state in the $^{76}$Ni core, which produces a multiplet of states with $J^{\pi}=3/2^--11/2^-$. The second is attributed to proton and neutron excitations across the shell gaps $Z=28$ and $N=50$, and results in the development of a deformed $K=1/2$ band with intrinsic quadrupole moment $Q_{0}\approx140$ efm$^2$. On average, the deformed states have 1.5 protons and 1.5 neutrons above the closed shells, compared with 0.5 protons and 0.5 neutrons for the spherical-like states. According to the calculations, the $E2$ transitions within the deformed band are expected to be much stronger than those between the spherical-like states and between the two structures.

In the previous work of Lokotko \textit{et al.} \cite{Lok20} on the lighter $\nu g_{9/2}$ odd-mass $^{A}$Co isotopes, excited spherical $(7/2^-_2)$ and $(9/2^-_1)$ levels arising from the coupling of the $\pi f_{7/2}^{-1}$ hole to the $(2^+_1)$ state in their $^{A+1}$Ni cores were identified. These are shown in black on the left panel of Fig. \ref{fig6}, together with the $7/2^-_2$ and $9/2^-_1$ levels calculated with the PFSDG-U interaction, depicted in red for the sake of clarity. At first sight, one can notice that the excitation energies of the $(7/2^-_2)$ candidates are on average 200 keV higher than their theoretical counterparts. If this systematic behaviour is extended to the lower-spin members of the multiplet in $^{75}$Co (shown in green in the figure), their excitation energies could very likely be degenerated with or slightly above the observed isomeric state. This shift would lend support to the interpretation of the isomer provided in Sec. \ref{sec:isomer}, which supports an $M3$ assignment for the 1914(2)-keV transition, corresponding to a $(1/2^-) \rightarrow (7/2^-)$ ground-state decay. 

Theoretically, the lowest-lying spherical-like state with $J^{\pi}=1/2^-$ arises from the particle-core coupling configuration $\pi f_{7/2}^{-1} \otimes 4^+_1(^{76}$Ni$)$. The question comes naturally: Is this state expected at about the excitation energy of the observed isomer? Considering that the experimental $(4^+_1)$ level in $^{76}$Ni lies at 1920 keV \cite{Sod15} and that the energies of the members of the $\pi f_{7/2}^{-1} \otimes 4^+_1(^{76}$Ni$)$ multiplet are expected to increase at decreasing spin \cite{Sat68,Paa73}, one can presume that the first spherical-like $1/2^-$ state lies above. Moreover, if the observed $(1/2^-)$ level is spherical, there still remains the question of why the deformed $1/2^-$ state, predicted at a much lower excitation energy, has not been observed as well. At least in the present data set, there are no indications for the existence of a low-spin $\beta$-decaying isomer from a least-squares fit of the ion-$\beta$ time correlation curve of $^{75}$Co. Based on these arguments, we presume that the $(1/2^-)$ isomeric level reported here for $^{75}$Co is more likely the bandhead of the deformed $K=1/2$ configuration. As it is shown in the right panel of Fig. \ref{fig6}, the deformed band would then be shifted up by $\sim$ 1 MeV with respect to the predictions of the PFSDG-U calculations. 
This tentative conclusion implies that the increasing trend of the prolate-deformed band towards $N=50$ in the odd-mass Co isotopes may be far more abrupt than expected by the PFSDG-U calculations (see the right panel of Fig. \ref{fig6}). This, in turn, points to a stronger dominance of spherical-like shapes at low excitation energies in the region immediately beneath $^{78}$Ni, posing the question of how fast deformation develops in the $N=50$ shell below $^{78}$Ni. Further spectroscopic data on this and more exotic $N\leq50$ nuclei will be necessary to provide answer to this question.

\section{Summary and conclusions}

The isomeric and $\beta$ decays of $^{75}$Co have been investigated at the RIBF facility at RIKEN (Japan) using the BigRIPS and EURICA setups. First spectroscopic information is provided for $^{75}$Co, for which a new isomeric transition at 1914(2) keV with a half-life of $t_{1/2}=13(6)$ $\upmu$s is reported. For the $\beta$-decay daughter $^{75}$Ni, new levels extending beyond those recently reported in Refs. \cite{Got20,Go20} are provided. In the case of the $\beta_n$ daughter $^{74}$Ni, the population of the $(8^+_1)$ candidate points to a similar feeding pattern as in the decay $^{73}$Co $\rightarrow$ $^{72}$Ni.

The nature of the observed states in $^{75}$Co and $^{75}$Ni has been discussed in terms of large-scale shell-model calculations using the PFSDG-U interaction in the $pf-sdg$ model space \cite{Now16}. In general, a good agreement between experimental and calculated results is found in $^{75}$Ni for excitation energies and log$ft$ values. In the case of $^{75}$Co, the observed isomeric state is proposed to have $J^{\pi}=(1/2^-)$, although a comparison with the PFSDG-U predictions reveals a 1-MeV discrepancy with the expected excitation energy of the prolate-deformed $J^{\pi}=1/2^-$ bandhead, leaving no clear interpretation for the nature of the observed state.

\section*{Acknowledgements}

The excellent work of the RIKEN accelerator staff for providing a stable and high-intensity beam of $^{238}$U is acknowledged. This work was partially supported by KAKENHI (Grants No. 25247045, No. 23.01752 and No. 25800130); U.S. DOE Grant No. DE-FG02-91ER-40609; Spanish Ministerio de Ciencia e Innovaci\'on Contracts FPA2009-13377-C02 and FPA2011-29854-C04, PID2019-104714GB-C21 (AEI), FPA2017-83946-C2-1-P (MCIU/AEI/FEDER), the Severo Ochoa Programme SEV-2016-0597 and grant PGC-2018-94583; Subvenciones grupos de excelencia de la Generalitat Valenciana: PROMETEO/2019/007; Spanish Comunidad de Madrid via the ``Atracci\'on de Talento Investigador'' Program No. 2019-T1/TIC-13194; European Regional Development Fund Contract No. GINOP-2.3.3-15-2016-00034; National Research, Development and Innovation Fund of Hungary via Project No. K128947; and the German BMBF Grant No. 05P19RDFN1. The authors acknowledge the EUROBALL Owners Committee for the loan of germanium detectors and the PreSpec Collaboration for the readout electronics of the cluster detectors. Part of the WAS3ABi was supported by the Rare Isotope Science Project which is funded by the Ministry of Education, Science and Technology (MEST) and National Research Foundation (NRF) of Korea.

\bibliographystyle{apsrev4-2}
\bibliography{bibliografia}

\end{document}